\bigskip
\documentclass[onecolumn,showpacs,preprintnumbers,amsmath,amssymb]{revtex4}
\usepackage{graphicx}
\usepackage{fancyhdr}
\usepackage{dcolumn}
\usepackage{bm}
\begin{document}
\preprint{ }
\title{Rearrangements of interacting Fermi liquids}
\author{Rong-Yao Yang  and  Wei-Zhou Jiang \footnotemark[1]}
\footnotetext[1]{ Email: wzjiang@seu.edu.cn}
\affiliation{Department of Physics, Southeast University,
Nanjing 210000, China}

{\begin{abstract} The stability condition of Landau Fermi liquid
theory may be broken  when the interaction between particles is
strong enough. In this case, the ground state is reconstructed to
have a particle distribution different from the Fermi-step function.
For specific instances, one case with the vector boson exchange
and another with the relativistic heavy-ion collision are taken into
consideration. With the vector boson exchange, we find that
the relative weak interaction strength can lead to the
ground-state rearrangement as long as the fermion mass is large
enough. It is found that the relativistic  heavy-ion collision
may also cause the ground-state rearrangement, affecting the statistics of
the collision system.
\end{abstract}}

\keywords{Landau Fermi liquid theory, ground-state instability,
relativistic heavy-ion collisions, vector bosons}

 \pacs{ 71.10.Hf, 24.10.Jv, 26.60.+c, 25.75.-q}

\maketitle \baselineskip 20.6pt

\section{Introduction}
Since the high $T_{c}$ superconductor and quantum Hall effect were
discovered in 1980s,  more and more systems which disobey the Landau Fermi liquid
theory have been springing up. It is known that the Landau Fermi liquid theory is
not valid for one-dimensional (1D) systems and strongly correlated 2D
liquids~\cite{LDG63,And90}. Khodel et al.~\cite{KVA90,ZMV99}
demonstrated that a state with the fermion condensation or
multi-connected momentum distribution arises as the stability
condition is broken and accordingly the ground state is rearranged.
When the ground-state rearrangement occurs, systems can possess novel
properties that are characteristic of a non-Fermi-liquid behavior.
Such a rearrangement can also occur in nuclear ground
states~\cite{Ll79,Ho81}, and can have possible implications in
astrophysical issues. For instance, the direct Urca process may be
remodelled when the rearrangement occurs, so that the fast cooling of
neutron stars is probable ~\cite{VDN00}.

In the present work, we consider two extreme examples that may
undergo a ground-state rearrangement due to repulsive interactions.
One is that the Fermi liquid system with the weak repulsion provided
by the vector boson exchanges. This may has a correspondence to the
aggregation of weakly interacting massive particles in compact stars.
In addition, it could exist the fifth force whose strength is
certainly very weak~\cite{fa89}. At least, this possibility has not
been excluded completely. Recently, considering this in terms of the
exchange of the U-boson, a few groups found that the U-boson that
couples with nucleons weakly can have a significant effect on
properties of neutron stars~\cite{kr09,WDH09,Zh11}. It is interesting
to examine whether the weak coupling can also cause the ground-state
rearrangement in specific systems. One may expect that the
rearrangement could have observable effects that may possibly be used
to constrain the new force.

Another example regards the furious deceleration in the relativistic
heavy-ion collisions in the time scale of the typical strong
interaction. According to the equivalence principle of the general
relativity, the deceleration or acceleration is identical to a local
gravitational field. As the usual particle statistics neglects the
space-time character, its consideration is of special interest and
significance~\cite{Eda11}. The present investigation serves the
purpose to  mimic fermion rearrangements in strongly curved space
that exist during spectacular supernova explosions and in the horizon
of a black hole. We will focus on determining the conditions of the
occurrence of the ground-state rearrangement in these two scenarios.

The paper is arranged as follows. In Sec. II, the stability condition
of ground states together with the ground-state rearrangement is
demonstrated for Fermi liquids. The numerical results are presented
in Sec. III. Finally, the summary is given.

\section{Stability condition of ground states}
\label{stability condition}
\subsection{Necessary stability condition for the Fermi liquid}
The ground-state stability condition can be established on the fact
that the variation  of the ground state energy is positive for any
admissible variations of distribution. Namely, the excitation states
have higher energy than the ground state, which means that
 \begin{equation}
 {\delta E_{0}} = \int[\epsilon(k)-\mu]\delta n(\textbf{k})\frac{d^{3}\textbf{k}}{(2\pi)^{3}},
 \label{eq1}
 \end{equation}
is positive when the conservative condition for the particle number
is satisfied: \(
\int{\delta}n(\textbf{k})\frac{d^{3}\textbf{k}}{(2\pi)^{3}}=0
\)~\cite{LD80}, where $\epsilon$ is the single-particle energy and
$\mu$ is the chemical potential.  The Landau Fermi liquid theory
tells us that the ground-state distribution of a homogeneous Fermi
liquid  is a Fermi-step function $n_{F}(k)=\theta(k_{F}-k)$ with
$k_{F}$ being the Fermi momentum. Namely, the states are filled below
the Fermi momentum and are empty above it. The stability condition is
satisfied if the single-particle energy above the Fermi surface is
larger than that below it. Otherwise, it is violated. That is to say,
 \[ s(k)=\frac{\epsilon(k)-\epsilon(k_{F})}{k-k_{F}}, \]
needs to be positive for the whole momentum range. In deed,
$\epsilon(k_F)$ is the chemical potential $\mu$ when the temperature $T \simeq 0$. As
$\lim_{k{\rightarrow}k_{F}}s(k)=\frac{d\epsilon(k)}{dk}\mid_{k=k_{F}}$,
the ground state of Landau Fermi liquid theory becomes unstable provided that
$d\epsilon/dk$ is  negative or zero in the vicinity of the Fermi
momentum. In this case, the mass of quasi-particles $M_Q=k_{F}/v_{F}$
is consistently negative or infinity once the value of the group
velocity $v_{F}\leq0(v_{F}=\frac{d\epsilon(k)}{dk}\mid_{k=k_{F}}$).

In a word, we obtain a necessary stability condition for the Landau
Fermi liquid theory: $d\epsilon/dk$ has to be positive near $k_{F}$
(In this paper we define the interval $0.95 k_{F} \sim 1.05 k_{F}$ as
the nearby $k_{F}$ to examine the stability condition).

\begin{figure}[thb]
\centering
\includegraphics[height=6.0cm,width=8.0cm]{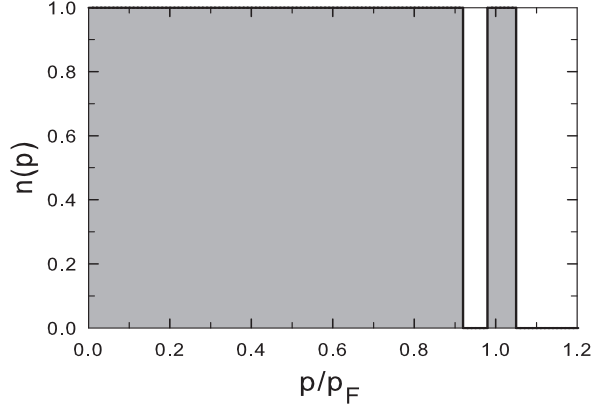}
\caption{ Reconstructed ground-state distribution after
the rearrangement. In this case, the ground state is
multi-connected.} \label{rearr}
\end{figure}

 With the breaking of stability condition, the ground state can only be
reconstructed by means  of the rearrangement. In Fig.~\ref{rearr}, we
show as an example the reconstructed ground-state distribution after
the rearrangement.

\subsection{Stability condition for systems with vector boson exchanges}
\label{vector bosons}  The first example that concerns  the
ground-state rearrangement is realized by the addition of repulsion
with the exchange of the vector boson. The addition of repulsion may
be supposed in strongly interacting nuclear matter or in weakly
interacting dark matter. We can examine the system stability by
analyzing its energy spectrum as above. In the relativistic
Hartree-Fock (RHF) approximation, the energy momentum distribution is
given as
\begin{eqnarray}
\epsilon(k)&=&\sqrt{k^{2}+M^{\ast2}}+\left[\frac{{\gamma}g_{\omega}^{2}}{2\pi^{2}m_{\omega}^{2}}\int_{0}^{\infty}
p^{2}n(p)dp+ \sum_{i=\omega,\rho}
\frac{g_{i}^{2}c_i}{8\pi^{2}k}\int_{0}^{\infty}pn(p)\ln\frac{(k+p)^{2}+m_{i}^{2}}{(k-p)^{2}
+m_{i}^{2}}dp \nonumber\right]\\
&& + \frac{{\gamma}g_{v}^{2}}{2\pi^{2}m_{v}^{2}}\int_{0}^{\infty}
p^{2}n(p)dp+\frac{g_{v}^{2}}{8\pi^{2}k}\int_{0}^{\infty}pn(p)\ln\frac{(k+p)^{2}+m_{v}^{2}}{(k-p)^{2}
+m_{v}^{2}}dp
\label{eq5}
\end{eqnarray}
where $M^{\ast}$ is the fermion effective mass which is density
dependent and moderately model-dependent, $m_{v}$ is the mass of the
vector boson of interest  (e.g., the U boson~\cite{Zh11}), $\gamma$
is the degree of degeneracy which is 4 with the spin and isospin
considered, $g_{v}$ is the boson-fermion coupling constant, and
$c_i=1,3$ for the $\omega$ and $\rho$ meson, respectively. The terms
in square brackets represent the contribution of the vector mesons
$\omega$ and $\rho$ in symmetric nuclear matter. In Eq.(\ref{eq5}),
besides the usual vector mesons $\omega$ and $\rho$ in the
relativistic nuclear models, we have included the additional vector
boson (denoted by the subscript $v$) in the original model to provide
the additional repulsion that may result in a limited modification to
the nucleon effective mass.

Substituting Fermi-step momentum distribution into Eq.(\ref{eq5}), we can obtain the explicit
expression of the derivative
\begin{eqnarray}
 \frac{d\epsilon(k)}{dk}&=&\frac{k}{\sqrt{k^{2}+M^{\ast2}}}-\sum_{i=\omega,\rho}\left(\frac{g_{i}^{2}c_i}{8\pi^{2}k}\int_{0}^{k_{F}}
p[\frac{1}{k}\ln\frac{(k+p)^{2}+m_{i}^{2}}{(k-p)^{2}+m_{i}^{2}}-\frac{2(k+p)}{(k+p)^{2}+m_{i}^{2}}+\frac{2(k
-p)}{(k-p)^{2}+m_{i}^{2}}]dp \nonumber\right)  \\
&& -\frac{g_{v}^{2}}{8\pi^{2}k}\int_{0}^{k_{F}}
p[\frac{1}{k}\ln\frac{(k+p)^{2}+m_{v}^{2}}{(k-p)^{2}+m_{v}^{2}}-\frac{2(k+p)}{(k+p)^{2}+m_{v}^{2}}+\frac{2(k
-p)}{(k-p)^{2}+m_{v}^{2}}]dp
\label{eq4}
\end{eqnarray}
It must be noted that the direct Hartree terms in Eq.(\ref{eq5}) have no contribution to the
derivative as they are independent of momentum $k$. As discussed before,
the momentum distribution of the ground state deviates from the step
function once $d\epsilon/dk\leq0$ in the vicinity of the Fermi
surface. When the coupling constant $g_{v}$ is larger than some
critical value $g_{vc}$, the zero point of $d\epsilon/dk$ will appear
near $k_{F}$. By numerical integration, we can search the zero point
for $d\epsilon/dk$ in the interval $0.95 k_{F}{\leq}k{\leq}1.05
k_{F}$, and the minimum $g_{v}$ for possessing at least one zero
point is the critical coupling constant $g_{vc}$.

\subsection{Necessary stability condition for collision systems}
\label{collision systems} It is well-known that the local  effect of
gravity on a physical system is identical with the effect of linear
acceleration of the system according to Einstein's equivalence
principle, and vice versa. In accelerated systems, the Hamiltonian
comprises a term provided by the non-inertial effect
\begin{equation}
H_{nin}=\frac{A^{2}}{c^{2}}\sum_{i}M_{i}z_{i}^{2}, \label{eq2}
\end{equation}
where $A$ is the acceleration of the system,  $c$ is the velocity of
light (equal to 1 in the natural unit), $M$ is the mass of particle
(usually the effective mass $M^{\ast}$ is used instead), z is the
particle's coordinate along the direction of acceleration of the
reference frame (in the case of collision, z equals to half the
distance between colliding particles, i.e. $z=r/2$ in the center of
mass reference frame)~\cite{BCA07}.

The non-inertial contribution in Eq.(\ref{eq2}) has a form of the
harmonic oscillator. In the Green function method or Feynman
diagrammatic approach, an explicit expression of the propagator of
the harmonic oscillator is mathematically complicated, and the
numerics seems to be not straightforward~\cite{Os98,Ma00}.  In order
to simplify this tedious issue, we suppose that the repulsive force
decelerating the particles during collision is provided by a
postulated vector meson exchange through a well-known Yukawa
coupling. At the same time, we assume that the step function
distributions are applied to systems prior to the collision. The
non-inertial effect can then be described by the vector meson
exchange  between nucleons which is sophisticatedly applied in modern
physics. Instead of the given non-inertial effect term, we can use
Yukawa potential $g^{2}e^{-rm_v}/4{\pi}r$, i.e., we
let $MA^{2}z^{2}=g^{2}e^{-rm_v}/4{\pi}r$, where $g$ is a dimensionless
coupling constant,$m_v$ is the mass of the postulated meson
 (generally in the same order of magnitude with nucleon mass),
 $r$ is the distance between nucleons. The propagator of Yukawa
potential is very simple: $F(k)=g^{2}/(k^{2}+m^{2}_v)$. The
single-particle energy for colliding systems can be expressed as
\begin{eqnarray}
\epsilon(\textbf{k})&=&\sqrt{\textbf{k}^{2}+M^{\ast2}}+\sum_{i=\omega,\rho}\left[
\frac{g_{i}^{2}c_i}{8\pi^{2}k}\int_{0}^{\infty}pn(p)\ln\frac{(k+p)^{2}+m_{i}^{2}}{(k-p)^{2}
+m_{i}^{2}}dp\right]\nonumber\\
 && +\int\frac{g^{2}}{(\textbf{k-p})^{2}
+m^{2}_v}n(\textbf{p})\frac{d^{3}\textbf{p}}{(2\pi)^{3}}
, \label{eq3}
\end{eqnarray}
where the direct Hartree terms are neglected because they do not
contribute to the momentum derivative that is used to determine the
critical coupling constant. Here, we quote the same coefficient $c_i$
as in Eq.(\ref{eq5}) by assuming the symmetric matter after the
collision. This is a reasonable assumption to neglect the effect of
isospin asymmetry because the coupling strength of the $\rho$ meson
is much less than that of the $\omega$ meson and the isospin
asymmetry is usually not large for the colliding system. After some
straightforward calculations, one can find that the derivative
$d\epsilon/dk$  is exactly the same as Eq.(\ref{eq4}), provided
 $g_{v}$ and $g_{vc}$ are replaced by $g$ and $g_{c}$ ,
respectively. With the given quantities $M^{\ast}$ and $m_v$ at the
density $\rho$, we adjust the coupling constant $g$ to exceed some
critical value $g_{c}$ where $d\epsilon/dk$ has the zero point in the
vicinity of the Fermi surface. Then we can find out the critical
coupling constant for the ground-state rearrangement.

\section{Numerical results and discussions}
\label{results}To obtain the fermion distribution function, one needs
first to solve the single-particle energy $\epsilon(k)$. Once the
rearrangement occurs, the fermion distribution function becomes
multi-connected and the numerical realization is quite tricky due to
the existence of sharp edges. We do not elaborate the numerical
details, and readers can be referred to Ref.~\cite{ZMV99} for
details. In the following, we discuss in turn the occurrence of
ground-state rearrangement in cases of the vector boson exchanges and
relativistic heavy-ion collisions.

\subsection{Rearrangement with vector boson exchanges}

Here we first  work on the relativistic model SLC~\cite{JWZ07}. The
additional vector boson is added to this model to provide additional
repulsion, and the nucleon effective mass is obtained in the RHF
approximation~\cite{NJW86}.  With the density-dependent nucleon
effective mass, we are able to investigate the rearrangement
self-consistently at various densities.  In what follows, we will
discuss the rearrangement in a simple system that consists of dark
matter in which the fermion mass is taken as a free parameter by
assuming that the dark matter is just weakly interacting. In this
case, we intend to see whether the rearrangement can occur with the
weak repulsion.
\begin{figure}[thb]
\centering
\includegraphics[height=8.0cm,width=8.0cm]{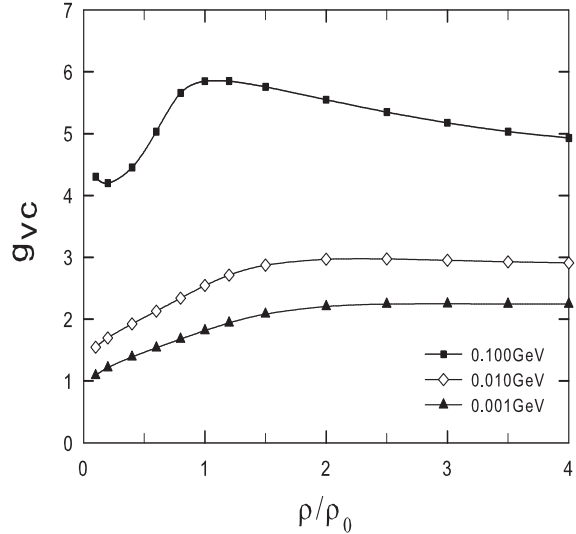}
\caption{The critical coupling constant as a function of density. It
is solved in the RHF approximation with the parametrization of the
model SLC with the addition of the additional vector boson exchange.}
\label{gc-rho}
\end{figure}

After obtaining the nucleon effective mass in the RHF approximation,
we can calculate the critical coupling constant for the ground-state
rearrangement.  In Fig.~\ref{gc-rho}, the critical coupling constant
is plotted as a function of density for various choices of the vector
boson mass: $m_{v} = 0.001, 0.010$ and $0.100 GeV$. As one can see
from Fig.~\ref{gc-rho}, the critical coupling constant $g_{vc}$
ascends monotonically with increasing the density when $m_{v}$ is
relative small ($m_{v} = 0.001,0.010 GeV$), while for the large mass
($m_{v} = 0.100 GeV$), $g_{vc}$ increases fast in the subsaturation
density region, then goes down slowly with the successive increase of
the density. In all cases, the critical coupling constant saturates
at high densities.

Recently, the vector boson beyond the standard model, dubbed the
U-boson that may accounts for the modification to the inverse square
law of the Newtonian gravity, has been introduced to compensate the
excessively small pressure suggested by the super-soft symmetry
energy~\cite{WDH09}. The super-soft equation of state (EOS) can be
successfully remedied by the repulsion provided by the U-boson with
the ratio of the coupling constant to the mass of vector boson
ranging from 7.07 $GeV^{-1}$ to 12.25 $GeV^{-1}$~\cite{WDH09}. Later
on, successive work indicates that the U-boson with the ratio around
10 $GeV^{-1}$ can even remedy largely the deviation between the soft
and stiff EOS's and the difference in the mass-radius relation of
neutron stars as well~\cite{Zh11}. We are intrigued to see whether
the ground-state rearrangement can occur with these ratio parameters
of the U-boson. Unfortunately, as displayed in the Fig.
~\ref{gc-rho}, various $g_{vc}/ m_{v}$ are all larger than 40
$GeV^{-1}$. Thus, it is usually impossible for nucleon systems to
rearrange its ground state with the regular ratio $g_{v}/ m_{v}$ that
is around 10 $GeV^{-1}$, and we are not able to anticipate the
constraint from whether or not the ground-state rearrangement occurs
on the U-boson parameters. On the other hand, this also reveals the
fact that using the Landau Fermi liquid theory is appropriate for the usual
nuclear Fermi liquid.

Now, we discuss dark matter that is invisible by the strong,
electro-weak but gravitational interaction. The properties of dark
matter are totally unknown to date~\cite{Be05}. For instance, the
mass of dark matter candidates can be assumed to vary from the
magnitude of eV to several hundred of GeV. Supposing that dense dark
matter can form, we are going to examine the ground-state
rearrangement of dark matter by taking the mass of fermionic dark
matter candidates as a free parameter. In this way, we still use
Eq.(\ref{eq4}) by simply neglecting the terms concerning the meson
exchanges that characterize the strong interactions. Specifically, we
calculate the $g_{vc}/m_{v}$ at $m_{v}=0.010 GeV$ for various $M$ at
the number density that equals to nuclear saturation density. The
results are listed in table ~\ref{tabthree}. It is seen that the
$g_{vc}/m_{v}$ drops clearly with increasing the mass of the dark
matter candidate. For $M=300.00 GeV$, the critical coupling constant
becomes as small as 0.11 which gives an interaction strength comparable to
the electromagnetic interaction. This reveals a
fact that the matter consisting of superheavy fermions can not be
regarded as a simple Fermi liquid as long as a weak repulsion
exists. Supposing that the heavy dark matter candidates stack up in
the core of dense stars, the rearrangement may easily occur therein
provided the repulsion comes up with the exchange of weakly
interacting bosons.

\begin{table}[bht]
\caption{Ratios of the critical coupling constant to the vector boson
mass at various $M$ for $m_{v}$=0.010 GeV and $\rho=0.16 fm^{-3}$
\label{tabthree}}
\begin{center}
\begin{tabular}{ c | c | c | c | c | c | c | c | c | c }
\hline\hline
$M(GeV)$ & 0.10 & 0.50 & 1.00 & 5.00 & 10.00 & 50.00 & 100.00 & 200.00 & 300.00 \\
\hline
$g_{vc}/m_{v}(GeV^{-1})$ & 352.44 & 248.15 & 183.29 & 83.27 & 58.91 & 26.35 & 18.63 & 13.18 & 10.76 \\
\hline
\hline
     \end{tabular}
     \end{center}
  \end{table}

\subsection{The case of collision}

For the heavy-ion collision, one needs in principle to determine the
density of crashed matter that depends on the collision energy.
However, it is complicated and beyond the scope of present work.
Considering the fact that the critical coupling constant for the
ground-state rearrangement does not change much as the density rises
up around the saturation density, see Fig.~\ref{gc-rho}, we
demonstrate the rearrangement at saturation density. For numerical
trials at other densities, the results are qualitatively similar, and
we will not specify the numerical details. Again, we adopt the basic
SLC parametrization to include the non-inertial effect and work out
the nucleon effective mass self-consistently in the RHF.

Though the non-inertial effect is hypothetically replaced by the
exchange of vector meson, we do not have the constraint on its mass.
In practical calculation, we thus change the postulated meson mass
from the lowest pion mass to the one close to 1 GeV. At the same
density, we can calculate different critical coupling constants for
different postulated meson masses. For instance of $m_v=0.5478 GeV$, we
obtain that at $k\simeq0.95 k_{F}$ the derivative $d\epsilon/dk$ is
negative for $g>g_{c}=31.91$. The results are tabulated in
Table~\ref{tabone}. We see that the critical coupling constant of the
postulated meson increases clearly with the mass, while the ratio of
the coupling constant to the mass changes much slowly.

\begin{table}[bht]
\caption{Critical coupling constants and accelerations for various
masses of a postulated vector meson. Here, we take $\rho=0.16
fm^{-3}$ and $R_0=10 fm$. \label{tabone}}
\begin{center}
\begin{tabular}{c | c | c | c | c | c | c | c | c}
\hline\hline
$m_v$ (GeV) & 0.1350 & 0.2626 & 0.4976 & 0.5478 & 0.6424 & 0.7800 & 0.8900 & 0.9578  \\
\hline
$g_{c}$ & 7.19 & 12.89 & 27.81 & 31.91 & 40.57 & 55.44 & 69.29 & 78.71 \\
\hline
$g_{c}/m_v$ (GeV$^{-1})$& 53.27 & 49.09 & 55.89 & 58.25 & 63.16 & 71.08 & 77.86   & 82.17  \\
 \hline
$A_{c}$(fm$^{-1}$) & 0.0441  & 0.0538 & 0.0855 & 0.0946 & 0.114 & 0.147 & 0.179 & 0.200 \\
\hline\hline
     \end{tabular}
    \end{center}
\end{table}

As one can anticipate, to
make sure the occurrence of rearrangement,  the inequality
${\mid}A{\mid} {\geq} A_{c}$ should hold, while $A_c$ is the minimum
acceleration to rearrange the ground state when the distance between
the colliding particles is given. Supposing the full stopping of the
colliding system leads eventually to a fireball with the radius of a
few femtometres, we estimate the $A_c$ by averaging the Yukawa-type
coupling and the coupling of the harmonic oscillator within such a
sphere, namely,
\begin{equation}\label{eq6}
A_{c}^2=\left(\int^{R_0}_{0} d^3rg_{c}^{2}\frac{e^{-rm}}{ r}\right)/{\int^{R_0}_{0} d^3r\pi
M^{\ast} r^2 }.
\end{equation}

In this way,   different critical decelerations for different
postulated meson masses can be estimated. Here, we use $\rho=0.16
fm^{-3}$ and $R_0=10fm$ as an example, and results are tabulated in
Table~\ref{tabone}. As one can see, for $m=0.5478 GeV$, with
$g_{c}=31.91$, we get ${\mid}A{\mid} \geq 0.0946 fm^{-1}$ for the
occurrence of rearrangement ( ${\mid}A_{c}\mid \simeq
8.50\times10^{30} m/s^{2}$). This critical deceleration is really
tremendous. However, it is not too difficult to reach such an
enormous deceleration in relativistic heavy-ion collisions. For
instance, two nuclei, being accelerated almost up to the light
velocity, crash against each other to be in full stopping in the time
scale of a few femtometres, e.g., $5-10 fm$, which is comparable to
the size of stopped matter. Then, the deceleration is $0.1-0.2
fm^{-1}$ which gives a reasonable and consistent magnitude of
deceleration with those in Table~\ref{tabone}. The present estimate
indicates that the non-inertial effect because of the huge
deceleration can lead to the ground-state rearrangement. Note that we
do not consider the effect of thermalization at the late stage of the
collision. The estimate also implies that as long as the
gravitational field is sufficiently strong, the ground-state
rearrangement may lead to deviation from the quasi-particle Fermi
liquid. For the case of larger density formed by the collision, we
can get moderately smaller critical decelerations. However, the
conclusion does not alter much for various densities of matter.

\section{Summary}
\label{summary}

We have investigated the applicability of the Landau Fermi liquid
theory and found that the necessary stability condition for this
theory may be broken in some circumstances. Once the necessary
stability condition is broken, the ground state is rearranged to be
different from the step function. In our relativistic model, the
nuclear Fermi liquid is stable with the usual strong interaction
strength and no rearrangement occurs. This is nevertheless in sharp
contrast with the system consisting of very heavy fermions that may
be regarded as dark matter candidates. We have found that it is
possible for heavy-fermion matter to rearrange the ground state even
by exchanging the weakly coupling vector boson. In the case of the
relativistic heavy-ion collisions, the ground state of the colliding
systems can be rearranged due to the strong non-inertial effect which
changes the ways of the quasi-particle occupation. This finding is
potentially instructive in investigating the fermion system during
the supernova explosions and in the horizon of black holes.

\section*{Acknowledgements}
The work was supported in part by the National Natural Science
Foundation of China under Grant Nos. 10975033 and 11275048 and the
China Jiangsu Provincial Natural Science Foundation under Grant
No.BK20131286.

\end{document}